# Web3D Graphics enabled through Sensor Networks for Cost-Effective Assessment and Management of Energy Efficiency in Buildings

Felix G. Hamza-Lup[1]
Armstrong State University
Savannah GA, USA

Marcel Maghiar [2]
Georgia Southern University
Statesboro GA, USA

## Abstract

*The past decade has seen the advent of numerous building energy efficiency visualization and simulation systems; however, most of them rely on theoretical thermal models to suggest building structural design for new constructions and modifications for existing ones. Sustainable methods of construction have made tremendous progress. The example of the German Energy-Plus-House technology uses a combination of (almost) zero-carbon passive heating technologies. A web-enabled X3D visualization and simulation system coupled with a cost-effective set of temperature/humidity sensors can provide valuable insights into building design, materials and construction that can lead to significant energy savings and an improved thermal comfort for residents, resulting in superior building energy efficiency. A cost-effective hardware-software prototype system is proposed in this paper that can provide real-time data driven visualization or offline simulation of 3D thermal maps for residential and/or commercial buildings on the Web.*

**CR Categories**: H.5.2 User Interfaces: GUI; H.5.3 Group and Organization Interfaces: Web-based interaction; I.3.7 Three-Dimensional Graphics and Realism: Virtual Reality; I.3.8 Applications; I.6.8 Types of Simulation: Visual;

## 1 Introduction

Sustainable methods of construction have made tremendous progress in the recent decades. For example, the German Energy-Plus-House technology [Voss and Musall, 2012] uses a combination of almost zero-carbon passive heating technologies and easily adaptable renewable energy technologies in order to build homes which have a positive energy balance enabling the home-owners to sell surplus energy to the national electricity grids and earning additional income.

Unfortunately, thermally deficient western methods of construction hardly tap into the huge potential of applying energy-efficiency technology. Moreover, the economically fastest growing regions in the world are now found in the countries of the global south located mainly in tropical and sub-tropical climate zones which have entirely different requirements when it comes to designing energy-efficient and low-carbon houses.

[1]email: Felix.Hamza-Lup@armstrong.edu
[2]email: MMaghiar@georgiasouthern.edu

With the fast (and often unregulated) economic growth, energy consumption in those parts of the world rises exponentially. The construction sector in those countries adapted fast to the western methods of construction and are now faced with major hurdles in assessing building energy efficiency, specifically for commercial buildings.

This article builds on, and expands on the research presented in [Hamza-Lup et. al., 2015] as it illustrates an effective, web-based visualization and simulation system coupled with a cost-effective set of sensors (e.g. temperature, relative-humidity) that can provide valuable insights into building's design, materials and construction leading to significant energy savings and an improved thermal comfort.

The article is organized as follows: in Section 2, we provide a study of existing models for energy assessment and simulation and introduce basic terminology; Section 3 describes the wireless sensors system employed for thermal data collection and communication; Section 4 provides a brief overview on 3D visualization methods, X3D rendering choices and presents preliminary experimental results for residential buildings; We present scalability issues for large commercial buildings in Section 5 and introduce the view-dependent rendering, advantages and limitations. Section 6 highlights the user base for the system, followed by the system assessment and validation methodology in Section 7. Section 8 concludes with a detailed discussion on future expansions and the importance of the proposed system.

## 2 Related Work

Conventional models of thermal representations are in use by construction professionals and HVAC (Heating, Ventilation and Air Conditioning) engineers for many years and they rely on proprietary thermal analysis software products like: SolidWorks, Ansys Advantage, Ansys CFX and many others that support CAD-based models integration. One of the goals of this research is to propose a modeling representation independent of proprietary software packages and their integration solutions, a representation that relies on the X3D open standard (Web3D, 2016).

International standards for thermal comfort for indoor air temperature and humidity as incorporated in the Nicol et. al. [1995] lacks representation and measurement methods and interpretation. They deal mainly with thermal comfort and the perceived values of temperatures from the building occupants. In the US alone, National Institute of Building Sciences (NIBS) are proposing (through High Performance Building Council) baseline standards [NIBS, 2015] for thermal performance of building enclosures, ASHRAE 90.1-2010, with certain levels of high performances, measurement and verification for design and construction of enclosure assemblies. Peeters et. al. [2009] presented a set of values and scales for thermal comfort evaluation in residential buildings. The rooms in each residence are categorized in three groups: bedroom, bathroom and other. Many factors (including thermal adaptation) are considered when deciding on the thermal comfort scales. Zhai et. al. [2006], presents a review on the complex computational fluid dynamics (CFD) simulations used for the past 20 years to model the flow of air in buildings. However, in authors' view, all simulation systems presented are proprietary and too complex to be applied in conjunction with real-time data.

Ham et. al. [2014] presents a thermography-based method to visualize the actual thermal resistance and condensation problems in buildings in 3D while taking static occlusions into account. Their experimental results show the promise of supporting retrofit decision-making for as-is building conditions; however, the method is dealing only with converting surface temperature data obtained from an IR camera into 3D visualization of energy performance metrics and possible condensation problems and is not addressing further thermal comfort considerations for as-built buildings. From a visual representation point of view, Wong and Fan [2013] research concurs that the lack of interoperability could be a factor limiting the application of BIM (Building Information Modeling) in building design and needs to be considered earlier in the planning stage for design processes. Through this paper, we make an attempt to address the shortage from a thermal comfort simulation perspective and to create an independent visualization and simulation platform accessible through a simple web browser.

Furthermore, as California lays the groundwork for adoption of a Zero Net Energy (ZNE)-ready code by 2020, it has recently restored its Advanced Homes Program to help meet this goal. This is an interesting attempt to promote innovation and, in this sense, one essential change has been to uncouple the program's targets from code level (i.e. incentives tied to percent better than code) and concentrate on tools and methods to better tie program outcomes to actual home energy performance. This way, the program experience is demonstrating that support on percent "better-than-code" metrics, code-compliance modeling and related tools, may hinder adoption of more advanced technologies and practices, obstruct innovation, and lead to misleading incentives aligned to code rather than to actual home energy performance [Christie et. al., 2014].

Rijal et. al. [2014] have conducted a thermal comfort and occupant behavior survey in thirty living rooms during the hot and humid season in the Kanto region of Japan. Their results showed that the residents adapt to the hot and humid environments by increasing the air movement using behavioral adaptation such as window opening and fan use. The study lacks simulation of thermal transfer of any kind and relies heavily on the statistical analysis performed on respondents based on the perceptive skin moisture sensation. Another study performed by Pitts [2013] deals with transition spaces like entrance foyers, circulation zones, lift lobbies, stairways and atria, and thermal comfort experiences. It both reviews existing reported research into comfort in such spaces are presented. The outcome of this work suggested only opportunities to reduce environmental conditioning and therefore energy use in such spaces. Lee [2008] developed a novel sensor network powered by the artificial light in order to achieve wireless power transfer and wireless data communications for thermal comfort measurements to implement a comfort-optimal control strategy for an air conditioning system.

One of the goals of this research is the investigation of X3D ISO standard for interactive 3D thermal maps representation with real-time data acquired from the sensor networks. The sensors can be connected to the HVAC equipment actuators, energy-consumption assessment and management being realized through the feedback of the thermal map information in the referred space.

From another point-of-view, in a study done by Lee et. al. [2014], the impact of three newly developed dynamic clothing insulation models on the building simulation is quantitatively assessed using the detailed whole-building energy simulation program, EnergyPlus, version 6.0. The limitations on this particular study were that the new clothing insulation models on energy and comfort was performed only for one particular climate (Chicago, IL, USA) and one particular system type (conventional forced air system equipped with variable speed central AHU and VAV box with reheat coil in each zone). The authors propose a wireless sensor-based system and a polygonal X3D thermal representation that is independent of these factors. The proposed software-hardware system allows an interactive 3D visualization of real-time data, as well as offline 3D simulation (e.g. speed playback) of the data collected over large timeframes (e.g. weeks/months).

## 3 Data Acquisition and Communication

The thermal map (comprising temperature and humidity) of a building is dynamic and depends on various factors like: the building envelope heat transfer, cooling/heating sources, and the humidifier/dehumidifier systems employed. The first means of heat transfer between indoor and outdoor climates occurs through the building envelope via conduction, convection and radiation. The second important means of heat transfer occurs through air exchange (infiltration/exfiltration).

The Data Acquisition System (DAS) proposed is collecting and feeding temperature and humidity data into the X3D visualization/simulation in real-time or offline. The DAS was designed to correspond to the integration requirements of already existing Building Management Systems (BMS) and was developed on a 3-Tier architecture as illustrated in Figure 1:

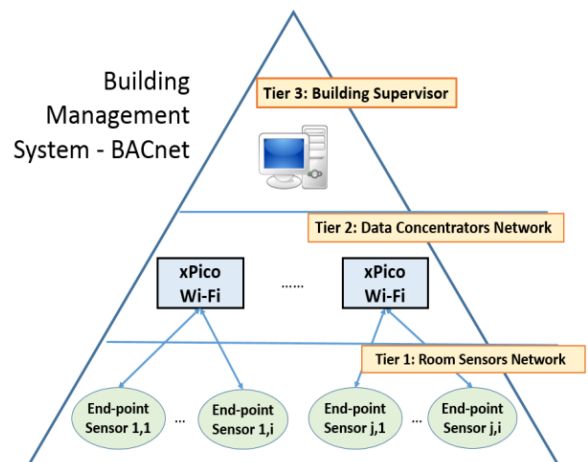

**Figure 1:** *Data Acquisition System: 3-Tiers*

- Tier 1: Room Sensor Network is a wireless network of smart sensors consisting of extremely low power consumption components. The sensors together with the associated circuitry for data transmission are denoted "End-points".

- Tier 2: Data Concentrators Network is represented by a low power consumption data concentrator at each room level, consisting of a data collection module from Tier 1 and a data transmission module to Tier 3
- Tier 3: The Building Supervisor is part of the BMS, and it is running over BACnet [Richard, 2013]. The Building Supervisor is responsible for the overall data collection and processing as well as data storage and on-demand retrieval.

Tier 1 relies on *m-bus* protocol over a Bluetooth 4.0 Low Energy (BLE) subnetwork [Bluetooth, 2014]. This communication protocol, register oriented, allows simple read/writes of the registers, data acquisition parameters control as well as potential control components for robotic actuators for automatic sensor distribution in the building. The network at Tier 2 is using communication protocols over an Ethernet TCP/IP network compliant with the 20 years old Building Automation and Control Network Protocol (BACnet) [NEWMAN, 2015]. Such protocols are compatible with HVAC communication and control and could be further expanded to manage these units.

### 3.1 Tier 1 - Room Sensors Network Design

To capture the heat transfer and relative humidity, the main component of the thermal data acquisition system is a set of low power consumption sensors integrated into a local Bluetooth network. The sensors have several features:
- Sensors dimensions and energy consumption is minimized;
- High accuracy for temperature .01°C, relative humidity ±2%
- High reliability (99.9%) of the network;
- Simple command and control communication protocol between the room server and end measurement sensor tunneled through Bluetooth wireless.
- Standard protocol BACnet [ANSI, 2010] between the room server and the building supervisor tunneled through Ethernet using Wi-Fi router as access point.

A cost-effective ($5/unit) set of temperature/humidity sensors SHT11 [Sensirion, 2015] (illustrated in Figure 2) with potential for mobility are deployed in each room. In case of large-scale deployment of a swarm of sensors (e.g. in case of modeling the thermal comfort for a commercial building) the total cost of the sensors system will be reduced by considerations of mass-production. As sensor placement strategy we choose to monitor the corners of each room, however other configurations may be chosen based on the architecture or other requirements (e.g. monitoring the HVAC vents).

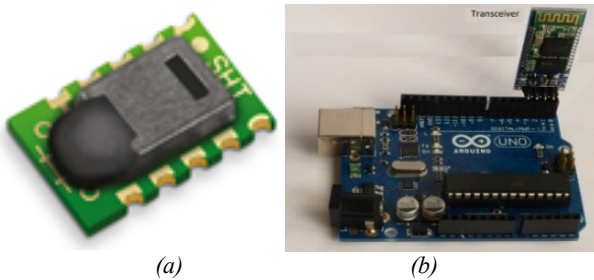

**Figure 2:** *(a) Sensirion SHT11 relative humidity sensor; (b) End-point sensor: SHT11, eXtreme Low Power PIC processor, Bluetooth Low Power transceiver (BLE)*

The sensors are light, accurate and have a short time constant. The power supply is a flat battery (1.8 Volts) that can power the system for several months (since the circuit has a power consumption of 3mW). The Block diagram in Figure 3 illustrates the connections on the End-point.

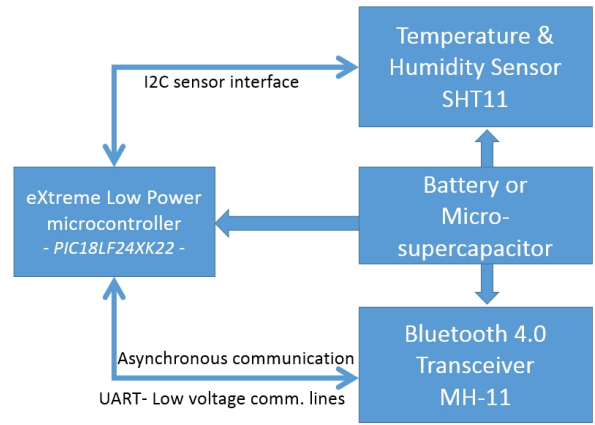

**Figure 3:** *Block diagram of the End-Point sensor system*

The microcontroller communicates with the sensor using the Inter-Integrated Circuit (I2C) interface and assures the synchronous sensor data acquisition. The microcontroller [PIC, 2015] receives two types of signals from the Data Concentrator (DC) from Tier 2: data synchronization signals and data interrogation signals, using the *m-bus* protocol. These signals can be extended to include mobility related signals in case the sensors are attached to robotic mobility systems for automatic movement and deployment.

While automatic sensor placement strategies are not explored in this research, such a possibility exist due to small sensor size. HexBugs [HEXBUG, 2015] carrying the small sensors represent a possible implementation based on the BEAM robotics concept established by TILDEN et. al. [1995]. The current sensor placement strategy is aimed at a uniform distribution of the sensors in the monitored volume and is highly dependent on the building geometrical features. Moreover after initial measurements and based on the requirements of the client we may choose to add additional sensors and/or cluster sensors in a problematic area, or areas where insulation problems have been detected. Tier 1 currently employs a star network topology.

Data communication between the sensors and the DC is based on the idea of circular memory buffers, physically located on the end-points. These buffers are written by the microcontroller based on the data collected from each sensor. Each line in the file contains information like the sensor id, the temperature value and the relative humidity value (unsigned integer). The buffer implements a FIFO strategy and it can hold up to 600 records. On each DC requests the most recent value is collected from the buffer. Consequently, by design, different query frequencies can be used by the DC. Frequencies of one query per second were explored and proved to be useful for short interval data collection (up to 1 day). Lower frequencies of one query per minute for medium interval data collection (up to 2 months) are appropriate.

The collected data is used in the color-coding scheme of the X3D thermal maps. The smart sensors network can be scaled on multiple levels and can include dozens of sensors in each room based on an id system and in accordance with the network topology desired. The data communication bandwidth can reach more than 1Mbps hence real-time data collection and processing is possible. We designed and implemented an 8 to 16 sensor-configuration per room depending on the accuracy of the temperature/humidity data requirements.

### 3.2 Tier 2 - Data Concentrators Network

The sensors communicate temperature and humidity measurements in an efficient manner to a local embedded server, implemented on a Lantronix (xPico Wi-Fi) [Lantronix, 2015] system, an ancillary microcontroller and the Bluetooth Low Power (BLE) master. Thus, from one side the DC will communicate and receive the data from the room sensors, and from the other side the DC will feed data into the building network through the Data Concentrators as illustrated in Figure 4.

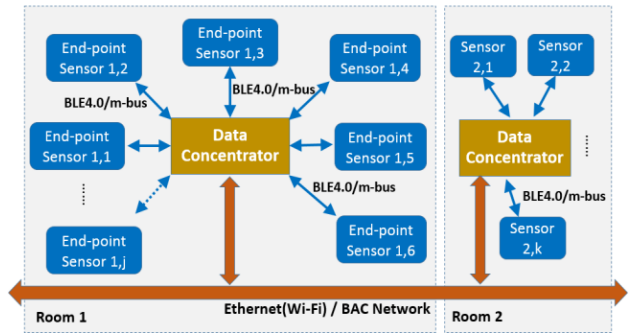

**Figure 4:** *Block diagram of Data Concentrators (DC)*

The Lantronix xPico Wi-Fi is an embedded access point system able to manage the request for data, issued by the BMS. The communication protocol between xPico and BMS is BACnet. At the same time the ancillary controller attached to xPico Wi-Fi assures the interrogation of all end-points (sensors) placed in the same room. Thus, in fact the DC plays also the role of a measurement "gateway". The DC network can also be exploited by connecting at the same access point a device like a laptop, tablet or smartphone, providing low-level sensor control. Such connections are very useful in the initial phase of setting the sensors and also for debugging purposes. Alternatively, sensor data can be saved on permanent storage for later use as required in the case of 3D simulation (i.e. playback of sensor data for extended periods of time - weeks or months with potential for predicted thermal pattern simulation). The topology used for tier-2 is a bus topology at each building level, implemented on the building backbone network.

## 4 Thermal Maps Visualization using X3D

Once the data has been collected it must be associated with the building structure. The building structural design can be easily obtained from various CAD-based modeling tools like SolidWorks [Solid Works, 2015] that allows the conversion to VRML format and hence to X3D. The complexity arises when one tries to visualize the volumetric representation of the temperature and the relative humidity in each room of the building based on the sensor data. 3D visualization and interactive navigation through the data sets are complex issues addressed next.

GPU's nowadays are powerful parallel processors that enable operations such as ray-casting at interactive rates. There are many techniques for volume rendering, some employ high performance computing methods others use cloud technology for large datasets and remote rendering [Bilodeau, 2012]. However we mostly focus on techniques for stand-alone workstations with standard graphics hardware as our system must be easily deployed in a web-based environment. Moreover, we do not require the users to have advanced hardware to run the application.

A common abstraction for a 3D visualization framework is a visualization pipeline [Moreland, 2013]. For such a processing pipeline, data representation and storage is essential for interactive speeds. Several well-known techniques can be employed as bricking (object space decomposition) [Fogal, 2013], multi-resolution hierarchies like octrees [Knoll, 2006], kd-trees [Fogal et. al., 2010] that allow efficient data traversal. Recently the sparse voxel octrees have gained attention in the graphics and gaming industry [Laine, 2010]. Several methods for rendering complex voxelized 3D models have been proposed [Heitz, 2012; Museth, 2013]. Data layout to efficiently access data on disk [Pascucci, 2002] as well as data compression [Rodriguez, 2014] have also been employed to improve rendering speed. While all these methods can significantly improve volumetric rendering, they are too complex and not available for deployment in a web environment at this time. We are proposing a non-volumetric (i.e. polygonal) representation for the thermal maps using the graphical primitives available in X3D.

Several X3D components have been investigated and their tradeoffs analyzed. For example the current X3D implementation offers a Fog node [X3D Fog, 2015] which allows linear or exponential visibility reduction as well as color addition in the environment simulating the humidity vapors in the air as illustrated in Figure 5; however, this implementation offers no control over the geometry and no possibility of color gradient. X3D geometric primitives (e.g. Box, Sphere) seem to be good candidates for 3D thermal map representation as their R, G, B and α-values (transparency) can provide an X3D thermal map as illustrated in Figure 5-b.

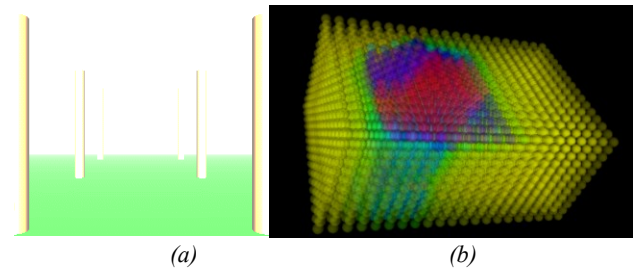

*(a)*  *(b)*

**Figure 5:** *Possible 3D Representations for temperature: (a) X3D Fog node, (b) X3D spheres with modified α-value.*

Since the X3D primitives' position and color/α-values can be controlled in a straight forward manner, the authors choose to model the volume in each room as a set of tangent semi-transparent color spheres. The sphere's color represents the average relative humidity (or temperature) at the respective location (as illustrated in Figure 5 - b). The α-value of the sphere is set in such a way that different levels of transparency can be achieved depending on the view-point distance from the building. Values are interpolated to determine intermediary relative humidity values. Linear interpolation is used currently as illustrated in Figure 6, however, other more complex models could enhance the visual representation of the sensor data (i.e. relative humidity/temperature values).

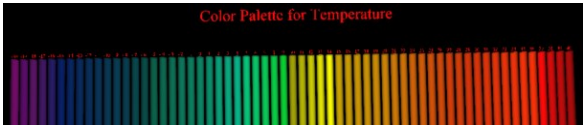

**Figure 6:** *Bell - shaped gradient colors*

## 4.1 Point Clouds versus X3D Primitives

The 3D thermal maps will require a very large number of polygons/vertices in the X3D representation. A large number of vertices will significantly reduce the interactivity of the X3D scene, making it difficult for the user to explore it at interactive speeds. The optimal framerate for interactive 3D visualization depends on many factors but primarily on the 3D visualization hardware. Moreover the Nyquist-Shannon sampling theorem [Teukolsky, 2007] implies that if motion in the X3D scene happens at a higher temporal frequency than half of the sampling rate (i.e. Nyquist frequency) the result will be temporal aliasing and visual artifacts generated by the high temporal frequencies interactions with a slow sampler. Hence, fast moving objects will not be represented in the X3D scenes.

For 2D visualization of the 3D scene on a regular LCD screen we observed that once the framerate drops below 10 frames per second (FPS) the user interactivity is compromised. We explored different X3D primitives (e.g. spheres, boxes, points) to obtain various 3D visualization models as illustrated in Figure 7.

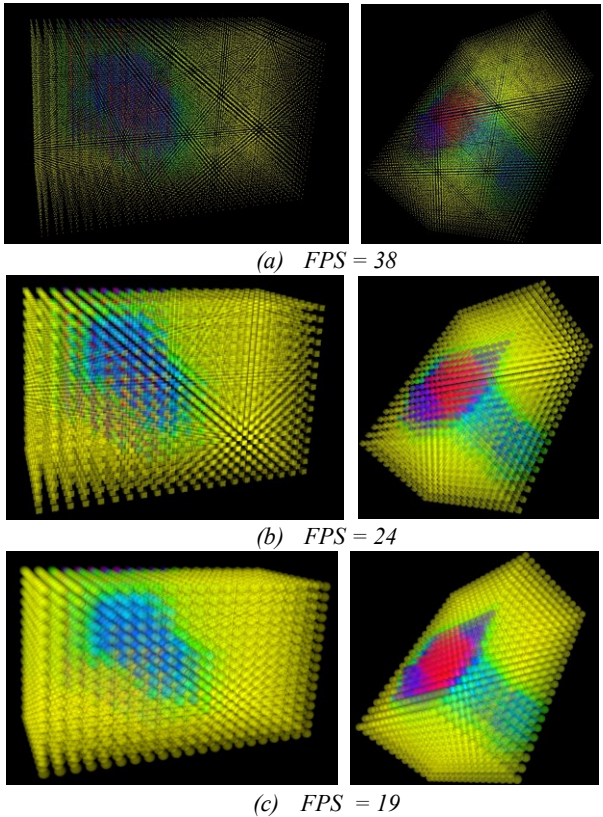

*(a)   FPS = 38*

*(b)   FPS = 24*

*(c)   FPS  = 19*

**Figure 7:** *(a) Point cloud rendering vs. (b) Box primitive vs. (c) Sphere primitive – illustrates hot spot in a room*

While a low polygonal count is best achieved through point-clouds for 2D projections of the 3D data, it is clear that X3D primitives are a better representation in spite of their slightly higher polygonal count.

Looking at the frame-rate, while point clouds offer the best performance, again the best visualization is observed using semitransparent primitives (e.g. Box, Sphere). Specifically for visualization on a 2D screen, 3D primitives have the advantage of providing substantial depth cues that are otherwise lost (e.g. in point cloud rendering).

## 4.2. Experimental Results - Residential Buildings

We configured a single family house with a set of thermal sensors and considered each rectangular room as a 3D container of semitransparent, tangent spheres that illustrates the 3D relative humidity map. The sensors have been placed in the corner of each room. The walls are rendered semi-transparent as illustrated in Figure 8-a, and as a wireframe, Figure 8-b. One can explore different viewpoints and visualize heat distribution in the house volume. As illustrated in this simulation example, the left corner of the building is overheated due to poor attic insulation or due to an insulation defect in the exterior walls.

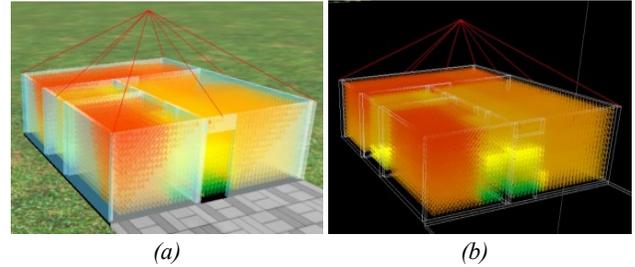

*(a)                                      (b)*

**Figure 8:** *Web-Browser view of residential building thermal maps: (a) flat, (b) wireframe.*

At current rates of 1Hz the sensor generated data can drive a one-frame per second real-time animation of the X3D thermal map using the X3D plugin for any browser. Of course higher framerates can be played back using collected data. Sensor data can be stored for large periods of time (e.g. months or years) and played-back as a sequence of thermal maps that change color in time, hence thermal map dynamicity can also be simulated at various time rates.

## 5. Scaling to Large Commercial Buildings

Scalable processing strategies for 3D visualization employ decomposition strategies both in the data domain and image domain and can employ single [Hadwiger et. al., 2012] and multi-pass [Hong et. al., 2005] rendering techniques. The X3D semi-transparent thermal map is scalable to large commercial buildings, however the difficulty in scaling resides in (1) the sensor placement strategy – a data domain issue, and (2) the X3D representation – an image-representation domain issue.

Time-based and scheduling strategies can also be employed for improved rendering speed. Since data from the sensors is stored and available in advance we can pre-process the data. The interpolation data for large time intervals is computed before rendering, offline.

For sensor placement we consider deploying a large set of sensors (sensor swarms) in the building using inexpensive miniature robots [Hoff, 2011] and a robot swarm algorithm for their even distribution. The automated sensor deployment and mobility system will be discussed elsewhere. Manual placing of sensors will be required for some rooms, specifically around thermally active elements (windows, HVAC units, etc.). As illustrated in Section 3, each room will consist of one DC and a set of sensors in a star network topology. The BMS can store large sets of data through an associated database (e.g. storing one year of sensor data is possible) and can also provide access to the collected sensor data through the existing wired or wireless network implemented in the building. As an example, a 14 sensors per room configuration will place sensors in each corner as well as in the middle of each wall (including ceiling and floor, considering a parallelepiped room structure). The sensor placement strategy is a complex issue that is not investigated here, however in many cases the location of the sensor will have a limited impact on the accuracy of the visualization as the data is collected from a large set of sensors over a large time-frame.

In terms of the X3D representation, for large commercial buildings, several polygon reduction techniques were investigated. First, we replaced the spheres due to their relatively high polygonal count, with the Contact Player [BitManagement 2015] tessellation approximately 300 polygons per sphere. Furthermore, we investigated the Box primitive (12 polygons), a custom made Tetrahedron (4 polygons) as well as simple Billboards (1-2 polygons). Average interactivity rates of 10 FPS were obtained with Tetrahedrons as well as with Billboards. Figure 9 illustrates the X3D thermal map for a six levels commercial building.

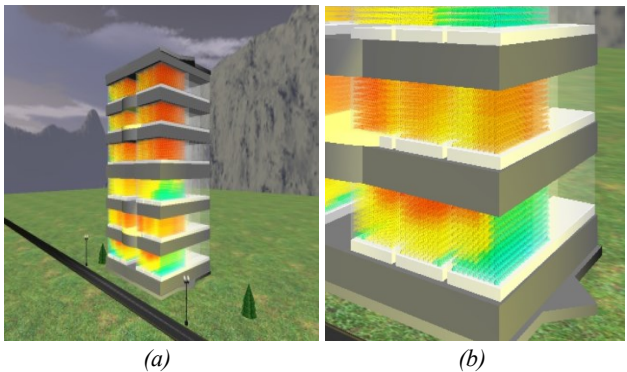

*(a)* *(b)*

**Figure 9:** *Web-browser view of commercial building thermal maps: (a) overall building envelope, (b) zoom on corner*

The data was artificially generated to replace the sensors' data since the authors focused on the 3D visualization issues and attempt at proving that the visualization can be easily scaled up to large commercial buildings and even city models.

### 5.1 View-Dependent Dynamic Rendering

To further improve the interactivity of the scene we have designed and implemented a view-dependent dynamic rendering algorithm in X3D that allows thermal map generation only for the regions of interest. The algorithm is based on [Yoon et. al., 2004] view-dependent algorithm for rendering at interactive speeds massive models. Figure 10 illustrates a user navigating in the X3D scene and approaching the floor level of the building (b) his/her viewpoint being approximatively one meter from the building corner. A fixed camera (view-point) located at 100 meters from the building, illustrated in (a) shows how the thermal map is generated dynamically only for the first level keeping the X3D model size and complexity relatively small as the user navigates through the scene.

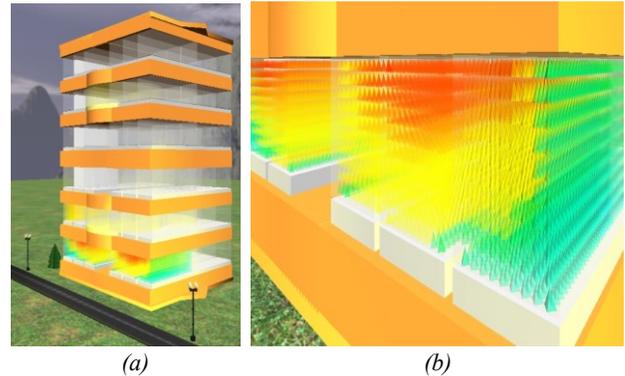

*(a)* *(b)*

**Figure 10:** *(a) Fixed camera showing the X3D scene; (b) user-view navigating through the scene, exploring the thermal map of the building's first level.*

Such view dependent optimizations will allow scaling of the visualization/simulation to very large commercial building and building complexes. Potentially the X3D scene could include several sections of a city, considering the data acquisition system is correctly deployed and integrated.

### 5.2 Advantages and Limitations

Scaling the model to various commercial buildings is especially valuable for both researchers and designers of HVAC systems by aiding in the pre-construction phase the evaluation of thermal comfort and indoor air quality. It also allows for designers to develop low-energy cooling and heating strategies such as natural ventilation systems and passive heating or cooling systems. The required data to analyze energy consumption in building is complex and should include marking data about the external environment, the shape and configuration of the building (building envelope), equipment loads, lighting, various mechanical systems and air distribution systems. Consequently, for accurate prediction of energy consumption, integrated visualization/simulation tools and sensors should be used.

Nevertheless, this study does not entail a Computational Fluid Dynamics (CFD) part to help with heat transfer in case of non-laminar air flow that may exist in some parts of the designed rooms and that is more computationally intensive. However, as Chung [2015] mentioned, while CFD have been utilized on building components or small-room-size analysis, whole-building energy analysis almost always is a simplified (time-stepped) heat-transfer analysis that does not use CFD. Users can produce results that are representative of the actual interior physics assuming a laminar air-flow, non-turbulent steady-state condition of airflow. Various simulations can be run assuming the locations and placements of the HVAC components throughout the rooms, lobbies and hallways. With the latest BIM developments, the researchers nowadays are able to use various modeling tools to illustrate the thermal loading, weather station data, shadow studies and CFD to dramatically explain the orientation's impact on the building envelope performance. However, visualizing X3D real-time data that may be collected through the proposed system for an existing building enables the building owner to make informed decisions regarding another building soon to be built and facing a certain direction. If the building is located in a high-solar radiation environment (like the desert) and the heat load on the building would have doubled the mechanical engineer's cooling requirements, the collected data and its simulation for BMS can potentially be calculated and graphically represented to show an owner the magnitude of realized savings for that particular design. Later in this process, changes can be applied in collaboration with the architect, when the cost of implementation for all necessary and/or desired changes is low compared with the overall budget allocated for the entire building.

On another note, the proposed X3D visualization system allows building level observations (e.g. roof, building envelope efficiency) as well as room level observations (efficiency of the HVAC system, cold/hot air drafts, window insulation efficiency, etc.). Moreover, these observations can be made for any time of the day and/or night when large variations of outdoor temperature are possible. Observations can also be made for large periods of time to monitor seasonal (spring, summer, autumn, winter) efficiency and thermal/humidity changes. Year-long observations are also possible in order to quantify the building ageing process and estimate energy losses. Many scenarios can be simulated, for example the effect of changing internal room separation or external architectural elements on the thermal maps, down to simple scenarios like forgetting a window opened in the building.

## 6 User Base

It is very essential to declare that the visualization system for user base is divided in several classes of potential users. Building Owners – who would like to discover potential issues with the degradation of the building insulation, building envelopes or building enclosures; they may want to address these issues to improve on future designs.

Energy Inspectors – energy audits are already required in EU [EU, 2012]. For example, specific measures and policies are to be implemented to ensure major energy savings for consumers and industry companies. Energy distributors or retail energy sales companies have to achieve 1.5% energy savings annually through the implementation of energy efficiency measures. Countries can opt to achieve the same level of savings through other means such as improving the efficiency of heating systems, installing double-glazed windows or insulating roofs. Also, the public sector in European countries should purchase energy efficient buildings, products and services every year, European governments will carry out energy efficient renovations on at least 3% of the buildings they own and occupy by floor area, empowering energy consumers to better manage consumption. In United States, only level-three energy audits by ASHRAE involves 3D computer models which would create simulations of real buildings, therefore changes to energy systems may be simulated with quite accurate results. Combining this process with construction-grade cost estimating enables to support informed investment decisions. These audits require access to energy consumption and cost data analysis for a duration of more than one year, typically to three years. Once the analysis is performed, recommendations can be made as to which improvements/upgrades are appropriate for the investment type required by a funding agency.

Resident Owners – they would like to know if there are problems with the building (house) in general. Major changes in temperature/humidity patterns can reveal structural damage (e.g. in areas with frequent earthquakes, high humidity or hot and humid climates). Also, designers, residential contractors and homeowners (occupants) are facing many energy related issues when planning and constructing new houses. Thermal issues may be revealed by using thermography studies (see Validation Methodology, Section 7), however thermal performance over a period of time through simulation of real-time energy usage may uncover important aspects of inefficient design for smaller spaces and volumes rather than in larger buildings. These design considerations may be taken into account during early planning. Cost-benefit evaluations can be made for houses in order to utilize an efficient energy plan over the life span of the construction or for retrofit of an existing equipment/material assembly. Thermal comfort is additionally important to be addressed for all human inhabitants through understanding how communication between volumes, interior-exterior environment and people may be visualized with energy simulations.

All building modeling visualization/simulations rely on proprietary visualization tools, available on the market for a fee (usually high-cost). Our implementation in X3D is not bound to a specific visualization tool, henceforth the user if free to choose any X3D player implementation solution. Furthermore, as X3D is an open international standard, many X3D player implementations are available and can be customized to fit the specific user's needs.

## 7 Validation Methodology using IR Imaging

A validation methodology for the 3D visualization and simulation system is proposed, using infrared (IR) thermography with a thermal imaging camera. The IR images are taken to expose areas on specific rooms with potential thermal issues that may have been started after construction (for as-built modeling), like poor or inadequate insulation, air leakage, heating and plumbing issues, water damages due to leaks, condensation, mold or identification and location of other potential problems that ultimately will affect thermal comfort.

Essential to both feedback practices in building design and performance of a constructed facility is the actual access to performance verification tools and data to inform construction contractors of installation quality, material issues or perhaps poor performance related to certain areas within the facility. In addition, technology solutions can be deployed to access and interpret smart energy meters data to verify performance or indicate the need for adjustments. Such validation approaches can reward good results and build up the confidence of building owners, lenders, and other business investors.

Building insulation quality becomes a pressing issue nowadays, as heating costs soar and heat leakage and thermal bridges are a most common phenomena. Thermal imaging is a powerful tool for determining the energy efficiency of spaces within buildings. Here it will be used for existing construction in case that potential issues are being investigated. The simulation part will confirm the dynamic nature of the issue, while the thermal camera will capture at a certain point in time the overall problem to be addressed. The procedure can save repair time and heating costs due to early issue detection and simulation study performed afterwards. Facility maintenance and facility managers can be involved in this process with the commitment to diagnose and address the issue immediately after discovery.

The validation method is implemented through a visual comparison of the thermal maps obtained through the IR camera and through the sensors at a specific moment in time. An example of such thermal maps is illustrated schematically in Figure 11.

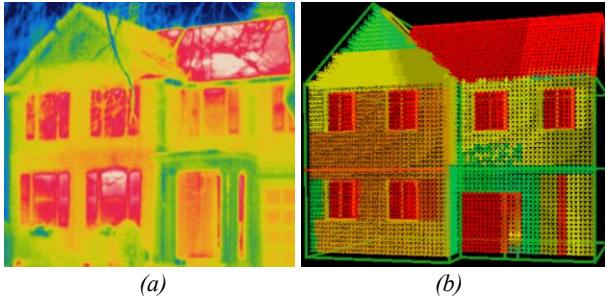

*(a)                           (b)*

**Figure 11:** *Validation method through visual comparison of: (a) IR camera imaging and (b) X3D thermal map*

It is assumed that the CAD or BIM models are available at this stage, in order to generate the X3D thermal map of the structure in focus.

In another example, the IR image of a potential problem on a water heater for a residential unit (Figure 12) can be validated through thermal imaging and discovered through X3D sensing and mapping presented earlier in this paper.

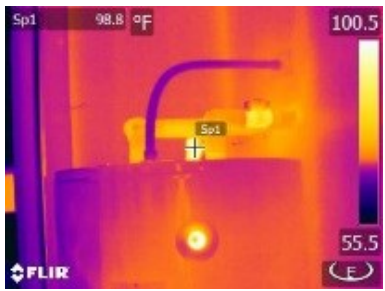

**Figure 12:** *Validation of a potential water heater problem through visual IR camera imaging*

Other various case scenarios may be investigated; the validation of a discovered issue from the thermal mapping within a building can be simply demonstrated with a thermal imaging camera which can later expose many other things invisible to the naked eye, like: patterns of heat loss, problems related to energy loss, missing or poor insulations, inefficient heating or cooling systems, water damage (Figure 13) or mold development in humid areas. These all can be part of energy audits reported to building owners and provided to them with exact measurements for localized temperatures and humidity values.

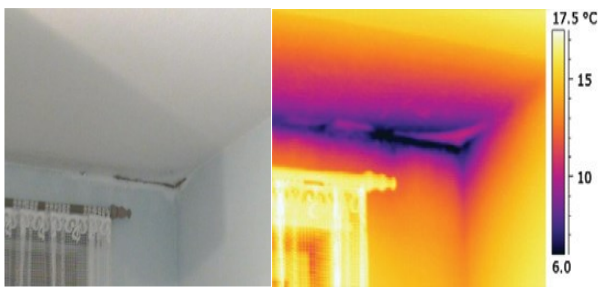

**Figure 13:** *Validation of a potential damage to a structure due to water leakage through visual IR imaging*

As a side-note, the discovered issue in Figure 13 on the corner of the room has potential to damage the structure due to water leakage. If the owner is not taking action and prescribe repair to be performed to the roof or to the exterior load-bearing walls of that building, the water damage might threaten the structural integrity of the entire building. It is noted how in this case the X3D mapping and validation process is used only prescriptive and for maintenance purposes only, taking into account the humidity and relatively low-temperature data acquired by end-point sensors.

## 8   Conclusions

This research presented a basic and cost-effective system for temperature/humidity data acquisition and an X3D module for 3D thermal maps visualization and simulation. The system can be used for assessing the thermal comfort for residential or commercial buildings. The X3D model coupled with the sensor system that we propose can also be employed in the design phase of commercial as well as residential buildings to improve the design of the building envelope and assemblies as well as to generate important energy savings. Another potential application is in the design and setup of the HVAC systems, specifically for commercial buildings where the thermal comfort has to be maintained under various conditions (e.g. large crowds and special relative humidity for storage rooms).

Building energy modeling can be applied early in the design development phase, as a collaborative effort between the energy consultant and the architect. Initial energy modeling scenarios may use forward simulation models such as this X3D computer simulation to predict approximate values for annual energy consumption and energy costs. The study will be expanded for case studies dealing with various human comfort zones during year-round seasons as humidity plays a major role in heat transfer. Many other factors influence human thermal comfort like: metabolic rate, clothing insulation, air temperature, mean radiant temperature and air speed. Psychological parameters such as individual expectations also affect thermal comfort, however this is beyond the scope of this paper.

A direction that the researchers will explore is the data correlation with external temperatures. The detailed analysis of temperature/humidity values recorded by sensors may be correlated with external temperature/humidity data received from weather stations in different locations/regions in the US, as they provide relatively accurate readings.

While Building Information Modeling (BIM) is not a new concept for the construction industry, the building industry has recently focused more consideration on the practical use of BIM and simulation tools. Even though the intelligence is lost due to transfer of these files (mainly parametric objects) into the X3D environment, the proposed system replaces the virtual model of a building with a three-dimensional virtual prototype containing a swarm of sensors that generates important thermal information ready to be used in the design, the assessment and the management process and/or operation of buildings. This allows energy consultants to convey embedded information to the design team, to integrate materials and specifications of the building into a database as the building is being designed, and to export input characteristics to other analysis tools, such as CFD or energy simulation for thermal comfort purposes.

One of the most exciting parts of this research is the cost-efficiency of the system to be implemented and the software-free web-based visualization of thermal maps. This process can be useful and can streamline the design and construction phases for HVAC components also. If different stakeholders in the project share a common web-based model that can be used for cost and energy analysis throughout the design and for comfort purposes, ulterior modifications, retrofitting of the HVAC components for comfort of the inhabitants can be eliminated and therefore saving labor and materials during occupancy phase. The proposed representation and modeling technique can also be distributed to the contractors and owners in order to improve the communication intent of the design. It is obvious that the sensor system can be used efficiently as a building diagnostic tool for potential issues that may appear after occupancy. Some of these issues may relate to the HVAC or other mechanical/electrical systems within the constructed facility.

On another note, the primary trade-off for the increased accuracy of hourly simulations is the increased time and effort that is required for making larger-scale buildings more interactive in order to illustrate the dynamic changes of the temperature. This graphical simulation should be used during the early planning phase to determine the order of magnitude of impacts related to design alternatives. The type of early design impact analysis allowed by X3D sensor-based graphical simulation may represent one of the key saving aspects during pre-construction, construction and operation phases and promote sustainable design of the respective residential or commercial building.

## References

ANSI. (2010) "ANSI/ASHRAE Addendum to ANSI/ASHRAE Standard 135-2008". BACnet ®, Approved by the ASHRAE Standards Committee on January 23, 2010.

BILODEAU B., SELLERS G., HILLESLAND K. (2012). "AMD GPU Technical Publications: Partially Resident Textures". In the Graphics Core Next, 2012.

BIT M. (2015). "BS Contact Player" Available online at: http://bitmanagement.com/. Accessed March 10, 2015.

BLUETOOTH (2014) "Bluetooth 4.0 BLE module - HM Bluetooth module datasheet V303", Texas Instruments.

CHRISTIE, M., ASPER, C., MORTON, J., BERRY, C. and BRAND, D. (2014). "Moving Beyond Better than Code: New Market Transforming Zero Net Energy Aligned Residential New Construction Programs". Proceedings of the 2014 ACEEE Summer Study on Energy Efficiency in Buildings. Washington, DC: ACEEE.

CHUNG, D. (2015). "Historic Building Façades: Simulation, Testing and Verification for Improved Energy Modeling". Journal of the National Institute of Building Sciences, Feb., (3), No. 1, pp.16-21.

EU (2012) "European Union Energy Efficiency Directive" Available online at: https://ec.europa.eu/energy/en/topics/energy-efficiency/energy-efficiency-directive. Accessed December 17, 2015.

FOGAL T., CHILDS H., SHANKAR S., KRÜGER J.,BERGERON R. D., HATCHER P. (2010) "Large Data Visualization on Distributed Memory Multi-GPU Clusters". In High Performance Graphics, pp. 57-66.

FOGAL T., SCHIEWE A., KRÜGER J. (2013). "An Analysis of Scalable GPU-Based Ray-Guided Volume Rendering". In IEEE Symposium on Large Data Analysis and Visualization, pp. 43–51.

HADWIGER M., BEYER J., JEONG W.-K., PFISTERH. (2012) "Interactive Volume Exploration of Petascale Microscopy Data Streams Using a Visualization-Driven Virtual Memory Approach". IEEE Transactions on Visualization and Computer Graphics 18 (12), pp. 2285–2294.

HAM, Y., and GOLPARVAR-FARD, M. (2014). "3D Visualization of thermal resistance and condensation problems using infrared thermography for building energy diagnostics." Visualization in Engineering, 2 (1), pp. 1-15.

HAMZA-LUP, F., Borza, P, Dragut, D. and Maghiar M. (2015) "X3D Sensor-based Thermal Maps for Residential and Commercial Buildings," Proceedings of the Web3D Conference, June 18-21, Heraklion, Crete, Greece, pp. 49-54.

HEITZ E., NEYRET F. (2012). "Representing Appearance and Pre-Filtering Subpixel Data in Sparse Voxel Octrees". In ACM SIGGRAPH / Eurographics conference on High-Performance Graphics, pp. 125–134.

HEXBUG (2015) "Scrab design reference from HexBugs", Available online at https://www.hexbug.com/mechanical/. Accessed December 15, 2015.

HOFF, N., WOOD, R., NAGPAL, R. (2011). "Effect of Sensor and Actuator Quality on Robot Swarm Algorithm Performance", International Conference on Intelligent Robots and Systems, Sept. 25-30, San Francisco, CA, pp. 4989 – 4994.

HONG W., FENG Q., KAUFMAN A. (2005) "GPU-Based Object-Order Ray-Casting for Large Datasets". In Eurographics/IEEE VGTC Workshop on Volume Graphics, pp. 177–24.

KNOLL A. (2006) "A Survey of Octree Volume Rendering Methods". GI Lecture Notes in Informatics. Proceedings of 1st IRTG Workshop, June 14-16, Dagstuhl, Germany.

LAINE S., KARRAS T. (2010). "Efficient Sparse Voxel Octrees Analysis, Extensions, and Implementation." NVIDIA Technical Report NVR-2010-001.

LANTRONIX (2015). "LANTRONIX xPicoWi-Fi Embedded Device Server". User Guide. Part Number 900-691-R Revision G, Lantronix, Inc.

LEE, D. (2008). "Development of Light Powered Sensor Networks for Thermal Comfort Measurement". *Sensors*. (8), pp. 6417-6432.

LEE, K.H., and SCHIAVON, S. (2014). "Influence of Three Dynamic Predictive Clothing Insulation Models on Building Energy Use", HVAC Sizing and Thermal Comfort. *Energies*. (7), pp. 1917-1934.

MORELAND K. (2013) "A Survey of Visualization Pipelines. IEEE Transactions on Visualization and Computer Graphics" 19 (3), pp. 367–78.

MUSETH K. (2013) "VDB: High-Resolution Sparse Volumes with Dynamic Topology". ACM Transactions on Graphics 32, (3), pp.1-22.

NEWMAN, H. M., (2015). "BACnet Celebrates 20 Years" ASHRAE Journal. 57 (6), pp. 36 - 42.

NIBS (2015). "National Performance Based Design Guide." Building Enclosures section. Available online at: http://npbdg.wbdg.org/enclosure.html. Accessed Mar. 5, 2015.

NICOL, F., HUMPHREYS, M., SYKES, O., and ROAF, S. (1995). "Standards for thermal comfort". TJ Press Ltd., UK

PASCUCCI V., FRANK R. J. (2002). "Hierarchical Indexing for Out-of-Core Access to Multi-Resolution Data". In Hierarchical and Geometrical Methods in Scientific Visualization. 2002, pp. 225–241.

PEETERS, L.F.R., DEAR, R. de, HENSEN, J.L.M. and D'HAESELEER, W. (2009). "Thermal comfort in residential buildings: Comfort values and scales for building energy simulation". Applied Energy, 86 (5), pp. 772-780.

PIC. (2015). "PIC18 (L) F2X/4XK22 – Data Sheet 28/40/44-Pin, Low Power, High-Performance Microcontrollers with XLP Technology", Microchip 2015, ISBN: 9781620763131.

PITTS, A. (2013). "Thermal Comfort in Transition Spaces." Buildings. 3 (1) , pp. 122-142.

RICHARD A. F. (2013) "BACnet Basics and Beyond", KMC Controls, Building Operating Management's NFMTVEGAS, National Facilities Management and Technology. Sept. 17-18, Mandalay Bay Convention Center.

RIJAL, H. B. (2014). "Investigation of Comfort Temperature and Occupant Behavior in Japanese Houses during the Hot and Humid Season". Buildings. 4 (3), pp. 437-452.

RODRÍGUEZ M., GOBBETTI E., GUITAN J., MAKHINYA M., MARTON F., PAJAROLA R., SUTER S. (2014). "State-of-the-Art in Compressed GPU-Based Direct Volume Rendering". Computer Graphics Forum 33 (6), pp.77–100.

SENSIRION (2015) "Sensirion Relative-Humidity Sensor". Specifications online at: http://www.sensirion.com/. Accessed March 10, 2015.

SOLIDWORKS. (2015). Available online at: http://www.solidworks.com. Accessed Sept. 10, 2015.

TEUKOLSKY, W.H, Vetterling, S.A, Flannery, B.P. (2007). "Section 13.11. Numerical Use of the Sampling Theorem", Numerical Recipes: The Art of Scientific Computing (3rd ed.), New York: Cambridge University Press, ISBN 978-0-521-88068-8.

TILDEN M. W, and BROSL HASSLACHER. (1995). "Robotics and Autonomous Machines: The Biology and Technology of Intelligent Autonomous Agents", LANL Paper ID: LA-UR-94-2636.

VOSS, K. and MUSALL, E. (2012). "Net zero energy buildings - International projects of carbon neutrality in buildings". 2$^{nd}$ edition, November 2012, Institut für Internationale Architektur-Dokumentation GmbH & Co. KG, München, ISBN 978-3-920034-80-5.

WONG, K. D., and FAN, Q. (2013). "Building information P5 modeling (BIM) for sustainable building design". Facilities, 31(3/4), pp. 138-157.

WEB3D (2016). "What is X3D". Web3D Consortium. Available online at: http://www.web3d.org/x3d/what-x3d. Accessed January 5, 2016.

X3D FOG. (2015). "The X3D Fog node". Available online at: http://doc.x3dom.org/developer/x3dom/nodeTypes/X3DFogNode.html. Accessed February 15, 2016.

YOON S-E, SALOMON, B., GAYLE, R., MANOCH D. (2004). "Quick-VDR: interactive view-dependent rendering of massive models", IEEE Visualization, pp.131-138.

ZHAI, Z. (2006). "Application of Computational Fluid Dynamics in Building Design: Aspects and Trends", Indoor and Built Environment, Aug. 2006, 15 (4), pp. 305-313.